\documentclass[12pt, a4paper]{article}
\usepackage[top=2.54cm, bottom= 2.54cm, left=2.54cm, right=2.54cm]{geometry}
\def\Dj{\hbox{D\kern-.73em\raise.30ex\hbox{-}
\raise-.30ex\hbox{}}}
\def\dj{\hbox{d\kern-.33em\raise.80ex\hbox{-}
\raise-.80ex\hbox{\kern-.40em}}}
\usepackage{epsfig, color}
\usepackage{amsmath,amsthm,amsfonts,amssymb,amscd,cite,caption}

\allowdisplaybreaks

\title{\Large \bf Quantitative Structure Property Analysis of Anti-Covid-19 Drugs}
\author{\bf Sunilkumar M. Hosamani \\[5mm]
{\it \normalsize Department of Mathematics, Rani Channamma University} \\[-1mm]
{\it \normalsize Belagavi-591156, India} \\[-1mm]
\normalsize E-mail: {\tt sunilkumar.rcu@gmail.com}\\
\normalsize\textbf{\textcolor[rgb]{1.00,0.00,0.00}{Dedicated to Corona-Warriors around the world}}
}
\date{}

\begin{document}
\maketitle
\vspace{-10mm}
\vspace{5mm}

\begin{abstract}
Inspired by recent work on anti-covid-19 drugs \cite{2} here we study the Quantitative-structure property relationships(QSPR) of phytochemicals screened against SARS-CoV-2 $3CL^{pro}$ with the help of topological indices like the first Zagreb index $M_{1}$, second Zagreb index $M_{2}$, Randi$\acute{c}$ index $R$, Balban index $J$ and sum-connectivity index $SCI(G)$. Our study has raveled that the  sum-connectivity index $(SCI)$ and the first Zagreb index $(M_{1})$ are two important parameters to predict the molecular weight and the topological polar surface area of phytochemicals respectively.
\\[3mm]
\\[3mm]
{\bf \AmS \; Subject Classification:} 05C90; 05C35; 05C12.
\end{abstract}

\baselineskip=0.30in

\section{Introduction}
According\textcolor{white}{i} to\textcolor{white}{i} the\textcolor{white}{i} World\textcolor{white}{i} Health\textcolor{white}{i} Organization(WHO)\textcolor{white}{i}, viral\textcolor{white}{i} diseases\textcolor{white}{i} continue\textcolor{white}{i} to\textcolor{white}{i} emerge\textcolor{white}{i} and\textcolor{white}{i} represent\textcolor{white}{i} a\textcolor{white}{i} serious\textcolor{white}{i} issue\textcolor{white}{i} to\textcolor{white}{i} public\textcolor{white}{i} health\textcolor{white}{i}. In\textcolor{white}{i} the\textcolor{white}{i} last\textcolor{white}{i} twenty\textcolor{white}{i} years\textcolor{white}{i}, several\textcolor{white}{i} viral\textcolor{white}{i} epidemics\textcolor{white}{i} such\textcolor{white}{i} as\textcolor{white}{i} the\textcolor{white}{i} severe\textcolor{white}{i} acute\textcolor{white}{i} respiratory\textcolor{white}{i} syndrome\textcolor{white}{i} coronavirus\textcolor{white}{i} (SARS-CoV) from\textcolor{white}{i} 2002 to 2003, and\textcolor{white}{i} H1N1 influenza\textcolor{white}{i} in 2009, have\textcolor{white}{i} been\textcolor{white}{i} recorded\textcolor{white}{i}.
The\textcolor{white}{i} coronavirus (COVID-19)\textcolor{white}{i} is\textcolor{white}{i} a\textcolor{white}{i} newly\textcolor{white}{i} emerged\textcolor{white}{i} human-infectious\textcolor{white}{i} coronavirus(CoV), pandemic\textcolor{white}{i} and\textcolor{white}{i} a\textcolor{white}{i}  global\textcolor{white}{i} health\textcolor{white}{i} emergency\textcolor{white}{i}. Unfortunately\textcolor{white}{i}, at\textcolor{white}{i} present\textcolor{white}{i} there\textcolor{white}{i} is\textcolor{white}{i} no\textcolor{white}{i} well-defined\textcolor{white}{i} treatment\textcolor{white}{i} or\textcolor{white}{i} therapeutics\textcolor{white}{i} against COVID-19\textcolor{white}{i} is\textcolor{white}{i} available\textcolor{white}{i} but\textcolor{white}{i} the\textcolor{white}{i} preventive\textcolor{white}{i} measures\textcolor{white}{i} are\textcolor{white}{i} being\textcolor{white}{i} recommended\textcolor{white}{i} worldwide\textcolor{white}{i}.
\vspace{2mm}

However\textcolor{white}{i}, the\textcolor{white}{i} clinical\textcolor{white}{i} trials\textcolor{white}{i} for\textcolor{white}{i} already\textcolor{white}{i} marketed\textcolor{white}{i} drugs\textcolor{white}{i} such\textcolor{white}{i} as\textcolor{white}{i} lopinavir, ritonavir, hydroxychloroquine, azithromycin, (Tirumalaraju \cite{20}) chloroquine (ClinicalTrials.gov, n.d.), Remdesivir (Tirumalaraju\cite{19}) etc. along\textcolor{white}{i} with\textcolor{white}{i} antibiotics\textcolor{white}{i} are\textcolor{white}{i} being\textcolor{white}{i} evaluated\textcolor{white}{i} to\textcolor{white}{i} treat\textcolor{white}{i} the\textcolor{white}{i} secondary\textcolor{white}{i} infections\textcolor{white}{i} (www.clinicaltrials.gov). All\textcolor{white}{i} of\textcolor{white}{i} the\textcolor{white}{i} drug\textcolor{white}{i} options\textcolor{white}{i} come\textcolor{white}{i} from\textcolor{white}{i} experience\textcolor{white}{i} treating\textcolor{white}{i} SARS\textcolor{white}{i}, MERS\textcolor{white}{i} or\textcolor{white}{i} some\textcolor{white}{i} other\textcolor{white}{i} new\textcolor{white}{i} influenza\textcolor{white}{i} virus\textcolor{white}{i} previously\textcolor{white}{i}. These\textcolor{white}{i} drugs\textcolor{white}{i} would\textcolor{white}{i} be\textcolor{white}{i} helpful\textcolor{white}{i} but\textcolor{white}{i} the\textcolor{white}{i} efficacy\textcolor{white}{i} needs\textcolor{white}{i} to\textcolor{white}{i} be\textcolor{white}{i} further\textcolor{white}{i} confirmed\textcolor{white}{i}. Few\textcolor{white}{i} COVID-19 vaccines\textcolor{white}{i} are\textcolor{white}{i} also\textcolor{white}{i} under\textcolor{white}{i} clinical\textcolor{white}{i} trials\textcolor{white}{i} such\textcolor{white}{i} as\textcolor{white}{i} Moderna's mRNA- 1273, first\textcolor{white}{i} US clinical\textcolor{white}{i} vaccine\textcolor{white}{i} funded\textcolor{white}{i} by\textcolor{white}{i} NIH's NIAID (National Institute of Allergy and Infectious Diseases) (Tirumalaraju \cite{18}). Thus, there\textcolor{white}{i} is\textcolor{white}{i} an\textcolor{white}{i} unmet\textcolor{white}{i} requirement\textcolor{white}{i} for\textcolor{white}{i} the\textcolor{white}{i} specific\textcolor{white}{i} anti-COVID-19 therapeutics\textcolor{white}{i} to\textcolor{white}{i} limit\textcolor{white}{i} the\textcolor{white}{i} severity\textcolor{white}{i} of\textcolor{white}{i} the\textcolor{white}{i} deadly\textcolor{white}{i} disease\textcolor{white}{i}.
\vspace{2mm}

Various clinicians and researchers are engaged in investigating and developing antivirals using different strategies combining experimental and in-silico
approaches see \cite{1,3,5,7,8,9,10,11,12,13,14,16,17,21,22,23,14}. The replication cycle of SARS-CoV-2 can be broadly divided into three processes – viral entry, viral RNA replication and lastly, viral assembly and exit from the host cell which is depicted in Figure 1.

\begin{center}
 \includegraphics[width = 5.0in]{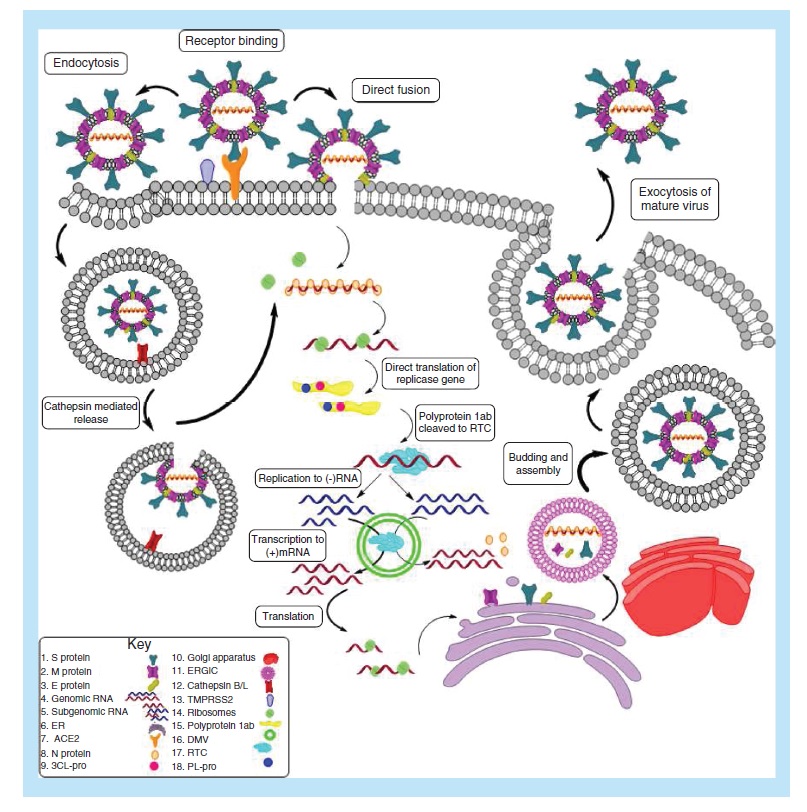}\\
 \textbf{Figure 1.} Replication cycle of SARS- COV-19.
\end{center}
Recent studies revealed that the genome sequence of SARS-CoV-2 is very similar to that of SARS-CoV. Recently, Qamar  et.al \cite{14} reported the following  phytochemicals screened against SARS-CoV-2 $3CL^{pro}$ which are depicted in Figure 2.
\begin{center}
 \includegraphics[width = 6.0in]{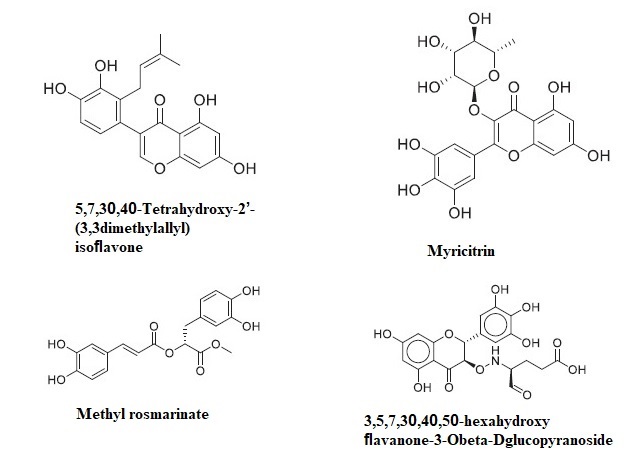}
 \end{center}
\begin{center}
 \includegraphics[width = 6.0in]{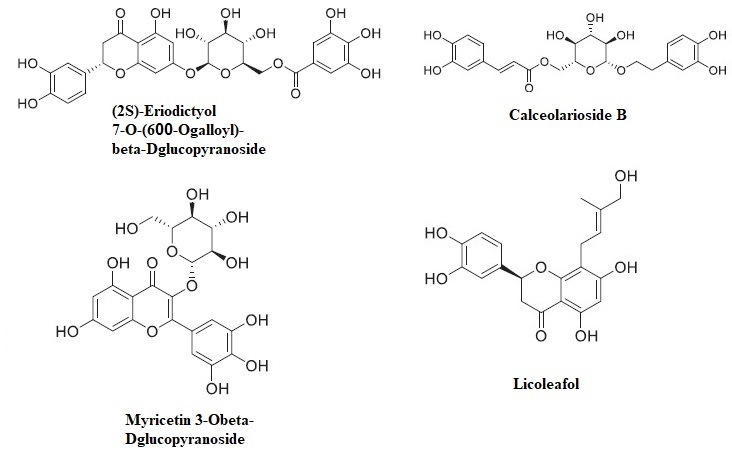}
\end{center}
\begin{center}
 \includegraphics[width = 6.0in]{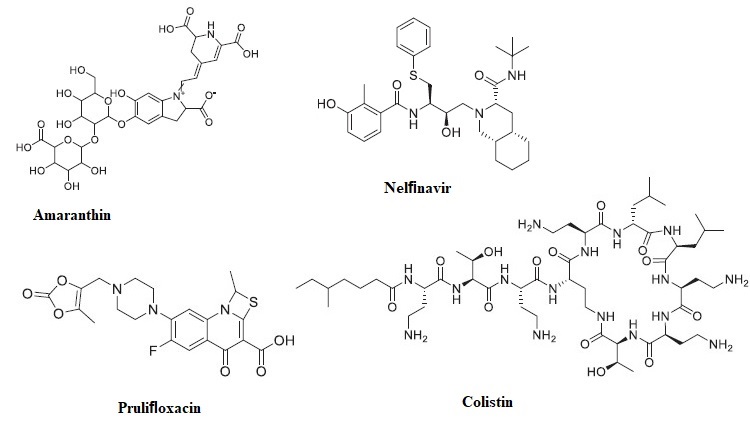}
\textbf{Figure 2.} Top ranked phytochemicals screened against SARS-CoV-2 $3CL^{pro}$.
\end{center}

\section{Molecular Graph and Topological Indices}

A\textcolor{white}{i} molecular\textcolor{white}{i} graph\textcolor{white}{i} is\textcolor{white}{i} a\textcolor{white}{i} connected\textcolor{white}{i} undirected\textcolor{white}{i} graph\textcolor{white}{i} corresponding\textcolor{white}{i} to\textcolor{white}{i} structural\textcolor{white}{i} formula\textcolor{white}{i} of\textcolor{white}{i} a\textcolor{white}{i} chemical\textcolor{white}{i} compound\textcolor{white}{i}, so\textcolor{white}{i} that\textcolor{white}{i} vertices\textcolor{white}{i} of\textcolor{white}{i} the\textcolor{white}{i} graph\textcolor{white}{i} correspond\textcolor{white}{i} to\textcolor{white}{i} atoms\textcolor{white}{i} of\textcolor{white}{i} the\textcolor{white}{i} molecule\textcolor{white}{i} and\textcolor{white}{i} edges\textcolor{white}{i} of\textcolor{white}{i} the\textcolor{white}{i} graph\textcolor{white}{i} correspond\textcolor{white}{i} to\textcolor{white}{i} the\textcolor{white}{i} bonds\textcolor{white}{i} between\textcolor{white}{i} these\textcolor{white}{i} atoms.\textcolor{white}{i} Molecular\textcolor{white}{i} graphs\textcolor{white}{i} have\textcolor{white}{i} fundamental\textcolor{white}{i} applications\textcolor{white}{i} in\textcolor{white}{i} chemoinformatics\textcolor{white}{i}, quantitative\textcolor{white}{i} structure-property\textcolor{white}{i} relationships(QSPR)\textcolor{white}{i}, quantitative\textcolor{white}{i} structure-activity\textcolor{white}{i} relationships(QSAR)\textcolor{white}{i}, virtual\textcolor{white}{i} screening\textcolor{white}{i} of\textcolor{white}{i} chemical\textcolor{white}{i} libraries,\textcolor{white}{i} and\textcolor{white}{i} computational\textcolor{white}{i} drug\textcolor{white}{i} design. QSPR\textcolor{white}{i}, QSAR\textcolor{white}{i} and\textcolor{white}{i} virtual\textcolor{white}{i} screening\textcolor{white}{i} are\textcolor{white}{i} based\textcolor{white}{i} on\textcolor{white}{i} the\textcolor{white}{i} structure-property\textcolor{white}{i} principle\textcolor{white}{i}, which\textcolor{white}{i} states\textcolor{white}{i} that\textcolor{white}{i} the\textcolor{white}{i} physicochemical\textcolor{white}{i} and\textcolor{white}{i} biological\textcolor{white}{i} properties\textcolor{white}{i} of\textcolor{white}{i} chemical\textcolor{white}{i} compounds\textcolor{white}{i} can\textcolor{white}{i} be\textcolor{white}{i} predicted\textcolor{white}{i} from\textcolor{white}{i} their\textcolor{white}{i} chemical\textcolor{white}{i} structure\textcolor{white}{i}. One\textcolor{white}{i} of\textcolor{white}{i} the\textcolor{white}{i} simplest\textcolor{white}{i} methods\textcolor{white}{i} that\textcolor{white}{i} have\textcolor{white}{i} been\textcolor{white}{i} devised\textcolor{white}{i} for\textcolor{white}{i} correlating\textcolor{white}{i} structures\textcolor{white}{i} with\textcolor{white}{i} biological\textcolor{white}{i} activities\textcolor{white}{i} or\textcolor{white}{i} physical-chemical\textcolor{white}{i} properties\textcolor{white}{i} involve\textcolor{white}{i} molecular\textcolor{white}{i} descriptors\textcolor{white}{i} called\textcolor{white}{i} topological\textcolor{white}{i} indices\textcolor{white}{i}.
\vspace{2mm}
The example of molecular graph is depicted in Figure 3.
\begin{center}
    \includegraphics[width = 1.8in]{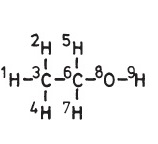}
    \includegraphics[width = 2.0in]{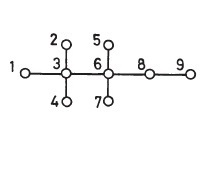}  \\
    \textbf{Figure 3.} Ethanole and its molecular graph.
\end{center}
Since physical properties or bioactivities are expressed in numbers whereas chemical structures are discrete graphs, in order to associate graphs with numbers one has to rely on graph-theoretical invariants such as local vertex invariants, e.g. vertex degree, distance sum, etc. Hundreds of topological indices have been introduced so far.
\vspace{2mm}

The main aim of this study is to develop a quantitative structure property relationship between two-dimensional(2D) topological indices, calculated physicochemical parameters of phytochemicals screened against SARS-CoV-2 $3CL^{pro}$. Experimental data used in this study were taken from \cite{14}. In this paper we have considered five topological indices viz., the first Zagreb index $M_{1}(G)$ \cite{6}, the second Zagreb index $M_{2}(G)$ \cite{6}, Randi$\acute{c}$ index $R(G)$ \cite{15}, Balban index $J(G)$ \cite{2,3} and the sum-connectivity index $SCI(G)$ \cite{25}. The formulae for these topological indices are give below:

\begin{eqnarray}
  M_{1}(G) &=& \sum\limits_{u \in V(G)} deg(u)^{2}
\end{eqnarray}

\begin{eqnarray}
  M_{2}(G) &=& \sum\limits_{uv \in E(G)} deg(u)\cdot deg(v)
\end{eqnarray}

\begin{eqnarray}
        R(G) &=& \sum\limits_{uv \in E(G)} \frac{1}{\sqrt{d_{G}(u) d_{G}(v)}}
 \end{eqnarray}

\begin{eqnarray}
        J(G) &=& \frac{m}{m+n-2}\sum\limits_{uv \in E(G)} \frac{1}{\sqrt{w(u) w(v)}}
 \end{eqnarray}
where $w(u)$ (resp. $w(v))$ denotes the sum of distances from $u$ (resp. $v$) to all the other vertices of $G$.

\begin{eqnarray}
  SCI(G) &=& \sum\limits_{uv \in E(G)} \frac{1}{d_{u}(G)+ d_{G}(v)}
\end{eqnarray}

The productivity of the above mentioned topological indices were tested using a data set of phytochemical, found at \cite{2} and https://pubchem.ncbi.nlm.nih.gov/. The data set consists of the following data: docking score, binding affinity, molecular weight and topological polar surface, which is given in Table 1.
\vspace{2mm}

\textbf{Table 1.} The physicochemical properties of phytochemicals.

\begin{tabular}{ccccccc}
  \hline
PubChem  & Phytochemical  &  Docking  & Binding  & Molecular  & Topological\\
IDs & Name &  score& affinity& Weight & Polar \\
&&(kcal/mol)&&& Surface\\
  \hline
11610052 & 5,7,30,40- Tetrahydroxy & -16.35 &	-29.57&	354.40&	107.00\\
&- 2'-(3,3-dimethylallyl) &&&& \\
&isoflavone & & & &\\
\hline
5281673 & Myricitrin & -15.64&	-22.13&	464.40&	207.00\\
\hline
6479915 & Methyl rosmarinate &-15.44&-20.62 &	374.30&	134.00\\
\hline
NPACT00105 &3,5,7,30,40,50- hexahydroxy &-14.42	&-19.10&.00&.00  \\
&flavanone-3-Obeta &  & & &\\
&- Dglucopyranoside & & & & \\
\hline
10930068 &(2S)-Eriodictyol &-14.41&	-19.47&	602.50&	253.00\\
& 7-O-(600-Ogalloyl)& & & &\\
& - beta-Dglucopyranoside& & & & \\
\hline
5273567 & Calceolarioside B&-14.36	&-19.87&478.40&186.00\\
&&&&&\\
\hline
5318606 &Myricetin & -13.70	&-18.42&480.40	&227.00 \\
&3-Obeta&  & & &\\
&- Dglucopyranoside& & & & \\
\hline
11111496 & Licoleafol&-13.63&-19.64&	372.40&	127.00\\
&& &&&\\
\hline
6123095 &Amaranthin &-12.67&	-18.14&	726.60&	346.00 \\
&&&&&\\
\hline
64143 & Nelfinavir & -12.20&	-17.31&	567.80&	127.00\\
\hline
65947 & Prulifloxacin &-11.32 &	-15.40&	461.50&	125.00 \\
\hline
5311054 & Colistin &-13.73&	-18.57&	1155.40&	491.00 \\
 \hline
\end{tabular}
\vspace{2mm}

\noindent \textbf{Note:} The molecular weight and topological polar surface of NPACT00105 could not find. Therefore, we do not include this molecule for QSPR-analysis.
\vspace{2mm}

The topological indices values of phytochemical structures ate given in Table 2.
\vspace{2mm}

\textbf{Table 2.} Topological indices calculated for phytochemicals used in the present study
\begin{tabular}{cccccc}
  \hline
 \textbf{Phytochemical Name} & &&\textbf{Molecular Descriptors}\\
 &&&&&\\
&  $M_{1}$& $M_{2}$& $R$ & $B$ & $SCI$ \\
  \hline
5,7,30,40- Tetrahydroxy & 138.0&	165.0 &	12.290601&	0.759930 &	12.738122\\
- 2'-(3,3-dimethylallyl) & & & & & \\
isoflavone & & & & &\\
\hline
Myricitrin &176.0 &	214.0 &	15.078295&	2.722760&	15.751722\\
\hline
Methyl rosmarinate & 128.0 &	144.0 &	12.256927	&3.478320&	12.485735\\
\hline
3,5,7,30,40,50- hexahydroxy  &166.0	&196.0&	15.116299&	3.542070	&15.551540\\
flavanone-3-Obeta& & & & &\\
- Dglucopyranoside & & & & &\\
\hline
(2S)-Eriodictyol & 228.0&	272.0&	19.904794&	2.295110&	20.815609\\
7-O-(600-Ogalloyl)& & & & &\\
- beta-Dglucopyranoside& & & & &\\
\hline
Calceolarioside B&172.0&	198.0&	15.150778&	2.765460&	16.604907\\
\hline
Myricetin  &182.0&	222.0&	15.488978&	3.206330&	16.159969\\
3-Obeta& & & & &\\
- Dglucopyranoside& & & & &\\
\hline
Licoleafol &142.0&	169.0&	12.811769&	3.442980&	13.248865\\
\hline
Amaranthin &300.0&	372.0&	25.102370&	2.118670&	26.436477\\
\hline
Nelfinavir &198.0&	228.0&	18.103175&	2.766410	&18.802009\\
\hline
Prulifloxacin &182.0&	224.0&	15.240091	&2.407980&	16.146148\\
\hline
Colistin &370.0&	413.0&	37.776411&	7.196940&	37.673916\\
 \hline
\end{tabular}

\subsection{Regression Models}
The following regression models have been used for the study:
\begin{itemize}
    \item Linear Model: $P = a (TI) + b$
  \item Quadratic Model : $P = a (TI)^{2} + b(TI) + c$
  \item Logarithmic Model: $P = a + b \ln(TI)$
  \item Multiple Linear Model: $P = a (TI_{1}) + b(TI_{2}) + c(TI_{3}) + d(TI_{4}) + e(TI_{5}) +f $
\end{itemize}
where $P$ is a physical property, $TI$ is the topological index, $a, b$ and $c$ are constants.\\

Next we present the regression models for docking score (DS) of phytochemical with the above mentioned topological indices.
\begin{description}
  \item[Linear Model:]
  \begin{eqnarray}
   \text{DS}  &=& 0.017 M_{1} -15.429\\
   \text{DS}  &=& 0.007 M_{2} -15.602\\
   \text{DS}  &=& 0.057 R -15.007\\
   \text{DS}  &=& 0.134 J -14.399\\
   \text{DS}  &=& 0.062 SCI -15.138
  \end{eqnarray}

  \item[Quadratic Model:]
  \begin{eqnarray}
  \text{DS}  &=& 0.070M_{1}^{2} -22.256\\
   \text{DS}  &=& 0.060 M_{2}^{2} -(9.501E-005)M_{2} -22.310\\
   \text{DS}  &=& 0.641 R^{2} -0.012R-21.064\\
   \text{DS} &=& 0.985 J^{2} -0.101J-15.841\\
   \text{DS} &=& 0.652 SCI^{2}-0.012SCI-21.383
   \end{eqnarray}

  \item[Logarithmic Model:]
  \begin{eqnarray}
  \text{DS}  &=& 1.951\ln(M_{1}) -24.213\\
   \text{DS}  &=& 0.057 \ln(M_{2}) -25.114\\
   \text{DS}  &=& 1.704 \ln(R) -18.519\\
   \text{DS}  &=& 0.822 \ln(J) -14.818\\
   \text{DS}  &=& 1.726 \ln(SCI) -18.93
\end{eqnarray}
\end{description}
\begin{center}
\includegraphics[width = 3.0in]{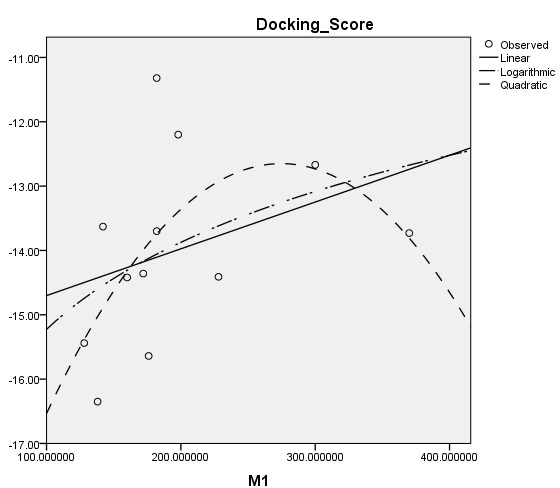}
\includegraphics[width = 3.0in]{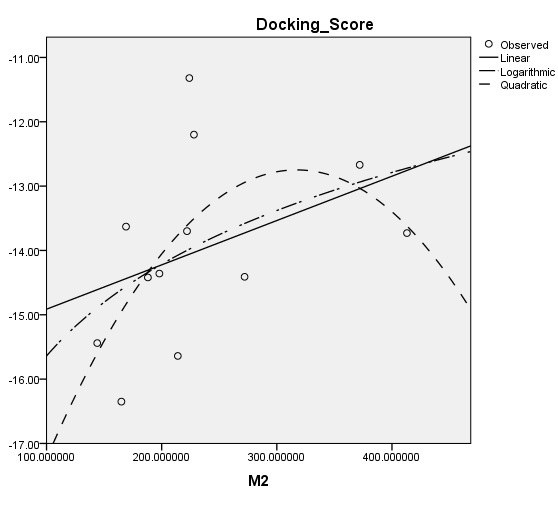}
\end{center}
\begin{center}
\includegraphics[width = 3.0in]{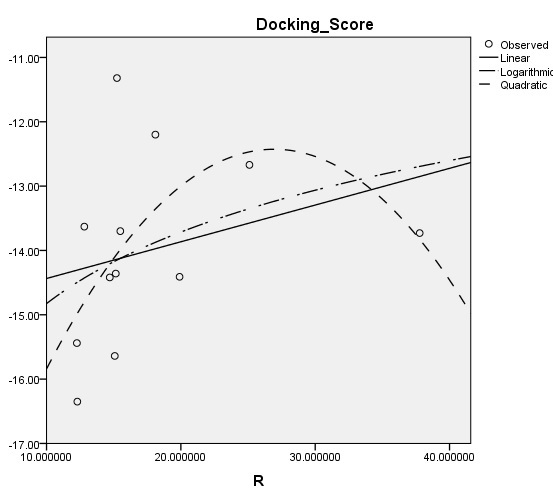}
\includegraphics[width = 3.0in]{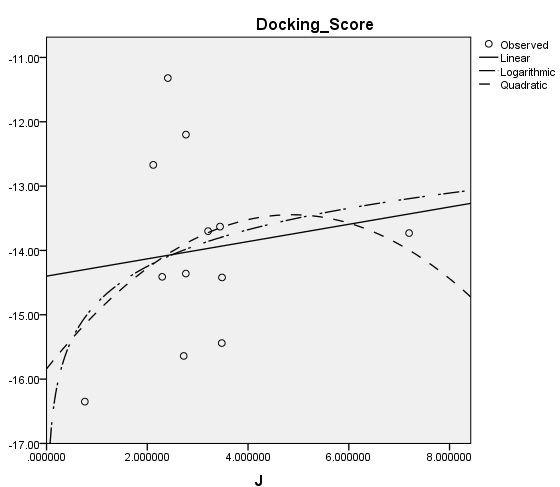}
\end{center}
\begin{center}
\includegraphics[width = 3.0in]{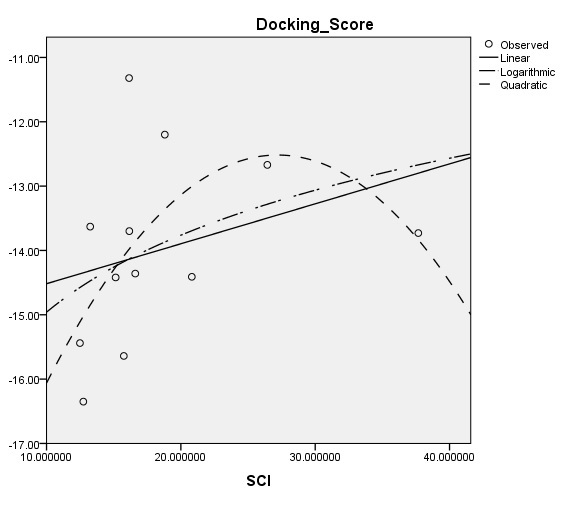}
\end{center}

\begin{center}
\textbf{Table 3.} Correlation coefficient, $F$ and $S$ values. \\
\begin{tabular}{|c|c|c|c|}
  \hline
  Model No & $R^{2}$ & $F$ & $S$ \\
\hline
Model 6&0.127&	1.449 & 0.256\\
Model 11 &0.177&	2.147& 0.174\\
Model 16&0.309&	2.011& 0.190\\
Model 7 & 0.151	&1.772 &0.213\\
Model 12& 0.198	&2.463 &0.148\\
Model 17 & 0.306	&1.984& 0.193\\
Model 8& 0.081 &	0.882 & 0.370\\
Model 13& 0.130	&1.500 &0.249\\
Model 18& 0.276& 	1.714&	0.234\\
Model 9 & 0.019	& 0.199& 0.665\\
Model 14& 0.084	& 0.915	& 0.361\\
Model 19& 0.081	& 0.394& 0.685\\
Model 10& 0.094&	1.037&	0.333\\
Model 15& 0.145	& 1.698& 0.222\\
Model 20&0 .288&	1.821& 0.217\\
\hline
\end{tabular}
\end{center}
Among all the topological indices used to predict docking score of phytochemicals, model 16 and model 17 were found to correlate well with correlation coefficient value $r= 0.309$ and $r =0.306$ respectively. In fact, the predicting power for topological indices considered here are too low for binding affinity of phytochemicals. Therefore, next we present the regression models for binding affinity (BA) of phytochemical with the above mentioned topological indices.
\begin{description}
  \item[Linear Model:]
  \begin{eqnarray}
   \text{BA}  &=& 1.607M_{1} -123.971\\
   \text{BA}  &=& 1.381 M_{2} -129.095\\
   \text{BA}  &=& 15.515 R -82.402\\
   \text{BA}  &=& 46.803 J +51.247\\
   \text{BA}  &=& 15.705 SCI -98.051
  \end{eqnarray}

  \item[Quadratic Model:]
  \begin{eqnarray}
  \text{BA}  &=& 0.273M_{1}^{2} + 0.003M_{2} +21.323\\
   \text{BA}  &=& -0.467 M_{2}^{2}+0.003M_{2} +103.085\\
   \text{BA}  &=& 19.238 R^{2} -0.076R-120.998\\
   \text{BA} &=& -75.217 J^{2} +14.519 J+258.086\\
   \text{BA} &=& 14.992 SCI^{2}+0.016SCI-89.554
   \end{eqnarray}

  \item[Logarithmic Model:]
  \begin{eqnarray}
  \text{BA}  &=& 4.591\ln(M_{1}) -43.904\\
   \text{BA}  &=& 353.161 \ln(M_{2}) -1656.138\\
   \text{BA}  &=& 339.535\ln(R) -764.752\\
   \text{BA}  &=& 99.187\ln(J) +94.221\\
   \text{BA}  &=& 346.307\ln(SCI) -797.856
\end{eqnarray}
\end{description}
\begin{center}
\includegraphics[width = 3.0in]{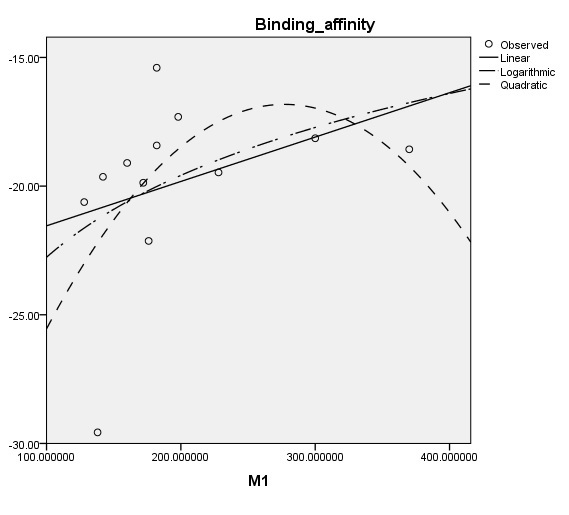}
\includegraphics[width = 3.0in]{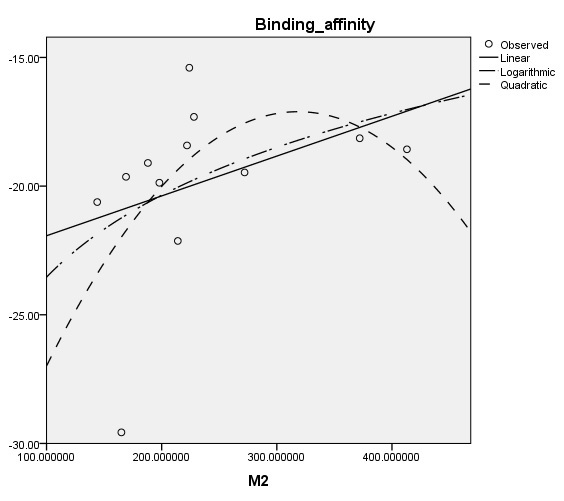}
\end{center}
\begin{center}
\includegraphics[width = 3.0in]{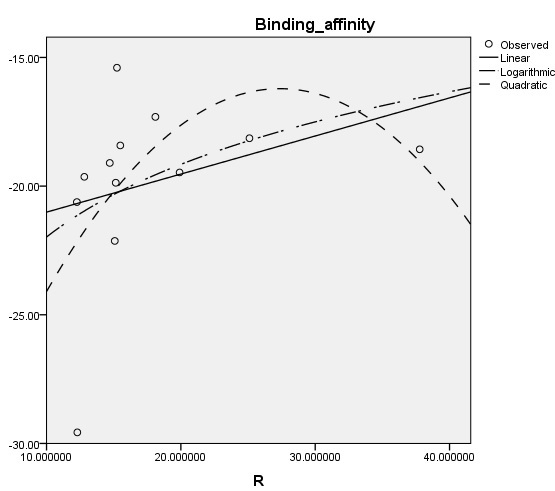}
\includegraphics[width = 3.0in]{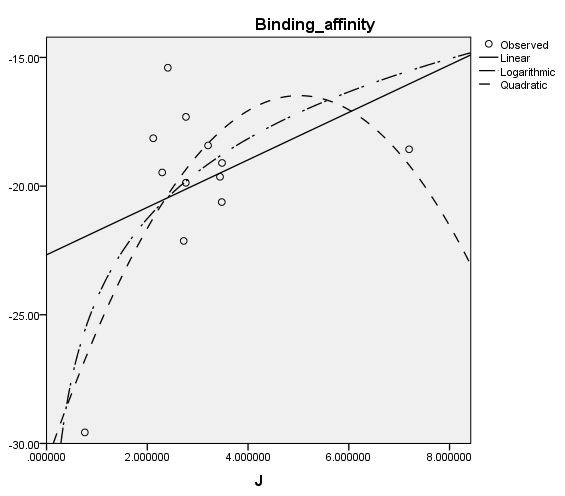}
\end{center}
\begin{center}
\includegraphics[width = 3.0in]{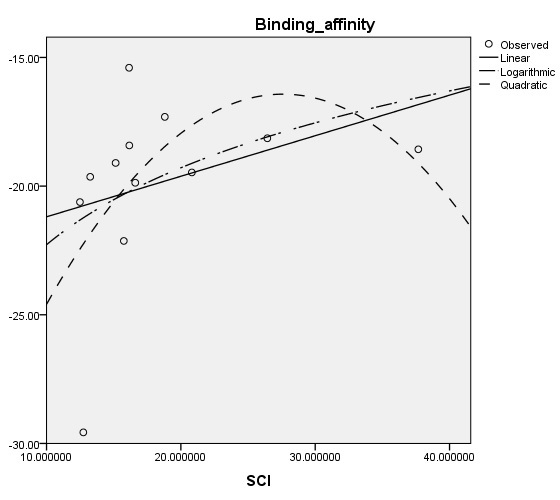}
\end{center}

\begin{center}
\textbf{Table 4.} Correlation coefficient, $F$ and $S$ values. \\
\begin{tabular}{|c|c|c|c|}
  \hline
  Model No & $R^{2}$ & $F$ & $S$ \\
\hline
Model 21&0.127&	1.449	&		0.256\\
Model 26&0.177&	2.147	&		0.174\\
Model 31&0.309&	2.011	&		0.190\\
Model 22&0.132&	1.519	&		0.246\\
Model 27&0.171&	2.061	&		0.182\\
Model 32&0.261&	1.585	&		0.257\\
Model 23&0.095&	1.046	&		0.331\\
Model 28&0.146&	1.710	&		0.220\\
Model 33&0.258&	1.561	&		0.262\\
Model 24&0.160&	1.899	&		0.198\\
Model 29&0.433&	7.635	&		0.020\\
Model 34&0.493&	4.374	&		0.047\\
Model 25&0.105&	1.174	&		0.304\\
Model 30&0.157&	1.866	&		0.202\\
Model 35&0.268&	1.650	&		0.245\\
\hline
\end{tabular}
\end{center}
Among all the topological indices used to predict binding affinity of phytochemicals, model 29 and model 34 were found to correlate well with correlation coefficient value $r= 0.433$ and $r =0.493$ respectively. In fact, the predicting power for topological indices considered here are too low for binding affinity of phytochemicals. Therefore, next we present the regression models for molecular weight (MW) of phytochemical with the above mentioned topological indices.
\begin{description}
  \item[Linear Model:]
  \begin{eqnarray}
   \text{MW}  &=& 3.323M_{1} -154.709\\
   \text{MW}  &=& 1.381 M_{2} -129.095\\
   \text{MW}  &=& 32.766 R -80.898\\
   \text{MW}  &=& 98.068 J+ 203.707\\
   \text{MW}  &=& 32.262 SCI -112.191
  \end{eqnarray}

  \item[Quadratic Model:]
  \begin{eqnarray}
  \text{MW}  &=& 0.358M_{1}^{2}+ 0.006M_{1} +168.241\\
   \text{MW}  &=& -0.467 M_{2}^{2}+ 0.003M_{2} +103.085\\
   \text{MW}  &=& 33.105 R^{2} -0.007R-84.420\\
   \text{MW} &=& -196.018 J^{2} +34.993J+ 702.219\\
   \text{MW} &=& 25.844 SCI^{2}+0.151SCI-33.663
   \end{eqnarray}

  \item[Logarithmic Model:]
  \begin{eqnarray}
  \text{MW}  &=& 730.933\ln(M_{1}) -3326.378\\
   \text{MW}  &=& 704.988 \ln(M_{2}) -3308.779\\
   \text{MW}  &=& 714.373 \ln(R) -1514.369\\
   \text{MW}  &=& 196.542\ln(J) +305.128\\
   \text{MW}  &=& 727.209\ln(SCI) -1581.401
\end{eqnarray}
\end{description}

\begin{center}
\includegraphics[width = 3.0in]{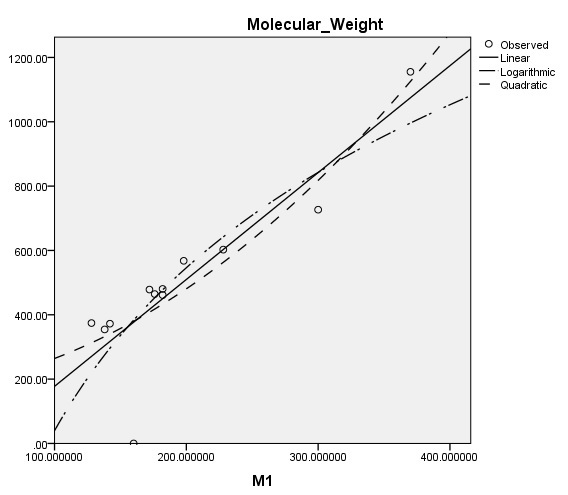}
\includegraphics[width = 3.0in]{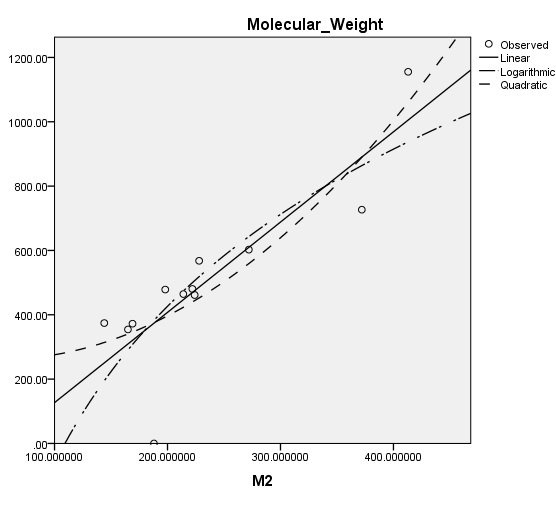}
\end{center}
\begin{center}
\includegraphics[width = 3.0in]{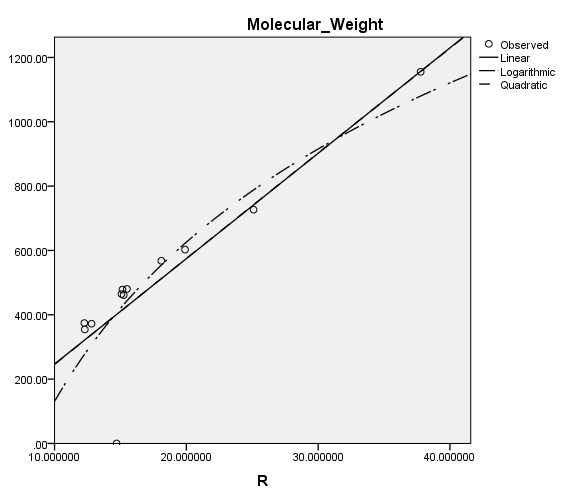}
\includegraphics[width = 3.0in]{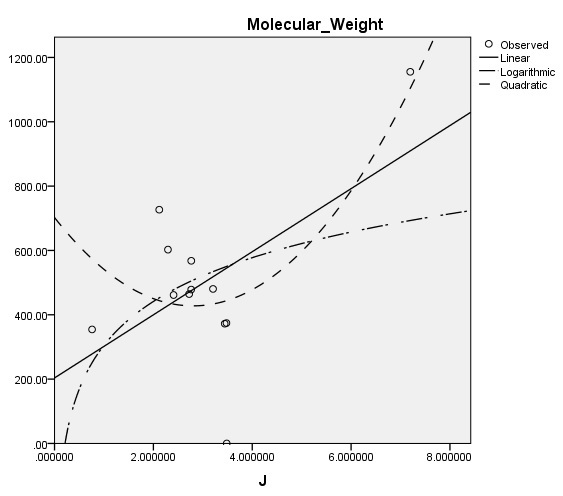}
\end{center}
\begin{center}
\includegraphics[width = 3.0in]{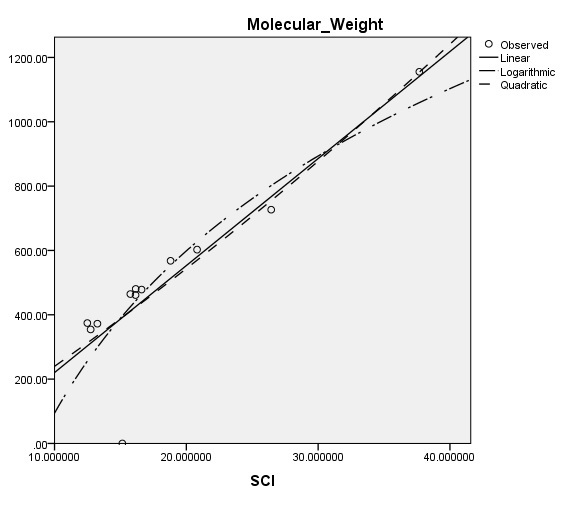}
\end{center}

\begin{center}
\textbf{Table 5.} Correlation coefficient, $F$ and $S$ values.\\ 
\begin{tabular}{|c|c|c|c|}
  \hline
  Model No & $R^{2}$ & $F$ & $S$ \\
\hline
Model 36&0.765&	32.588			&0.000\\
Model 41&0.718&	25.484			&0.001\\
Model 46&0.777	&15.680			&0.001\\
Model 37&0.723	&26.105			&0.000\\
Model 42&0.672&	20.489			&0.001\\
Model 47&0.742	&12.912			&0.002\\
Model 38&0.780	&35.410			&0.000\\
Model 43&0.761&	31.783			&0.000\\
Model 48&0.781	&16.030			&0.001\\
Model 39&0.308	&4.448			&0.061\\
Model 44&0.159	&1.887			&0.200\\
Model 49&0.472	&4.016			&0.057\\
Model 40&0.791	&37.849			&0.000\\
Model 45&0.761	&31.785			&0.000\\
Model 50&0.791&	17.037			&0.001\\
\hline
\end{tabular}
\end{center}
By looking at the above table we can see that the predicting power of above mentioned topological indices are good  with respect to molecular weight of phytochemicals. The correlation coefficient of the first Zagreb index ($M_{1}$) lies between 0.718 to 0.777, whereas, the range for the Zagreb index($M_{2}$) is lies between 0.672 to 0.742. For Randi$\acute{c}$($R$) index the $r$ values is lies between 0.761 to 0.781 and for the Balban index $J$ the $r$ value ranging from 0.159 to 0.472. Finally for sum-connectivity index, the $r$ value lies between 0.761 to 0.791. Except, the Balban index, all $TI's$ are shows good correlation coefficient. Among all $TI's$, the sum-connectivity index $SCI$ is a good candidate for predicting molecular weight of phytochemicals.

Next we present the regression models for topological polar surface (TPA) of phytochemical with the above mentioned topological indices.
\begin{description}
  \item[Linear Model:]
  \begin{eqnarray}
   \text{TPA}  &=& 0.017 M_{1} -23.268\\
   \text{TPA}  &=& 0.007 M_{2} -15.602\\
   \text{TPA}  &=& 0.057 R -15.007\\
   \text{TPA}  &=& 0.134 J -14.399\\
   \text{TPA}  &=& 0.062 SCI -15.138
  \end{eqnarray}

  \item[Quadratic Model:]
  \begin{eqnarray}
  \text{TPA}  &=& 0.154M_{1}^{2} -38.178\\
   \text{TPA}  &=& 0.060 M_{2}^{2} -(9.501E-005)M_{2} -22.310\\
   \text{TPA}  &=& 0.641 R^{2} -0.012R-21.064\\
   \text{TPA} &=& 0.985 J^{2} -0.101bJ-15.841\\
   \text{TPA} &=& 0.652 SCI^{2}-0.012SCI-21.383
   \end{eqnarray}

  \item[Logarithmic Model:]
  \begin{eqnarray}
  \text{TPA}  &=& 4.591\ln(M_{1}) -43.904\\
   \text{TPA}  &=& 0.057 \ln(M_{2}) -25.114\\
   \text{TPA}  &=& 1.704 \ln(R) -18.519\\
   \text{TPA}  &=& 0.822 \ln(J) -14.818\\
   \text{TPA}  &=& 1.726 \ln(SCI) -18.93
\end{eqnarray}
\end{description}
\begin{center}
\includegraphics[width = 3.0in]{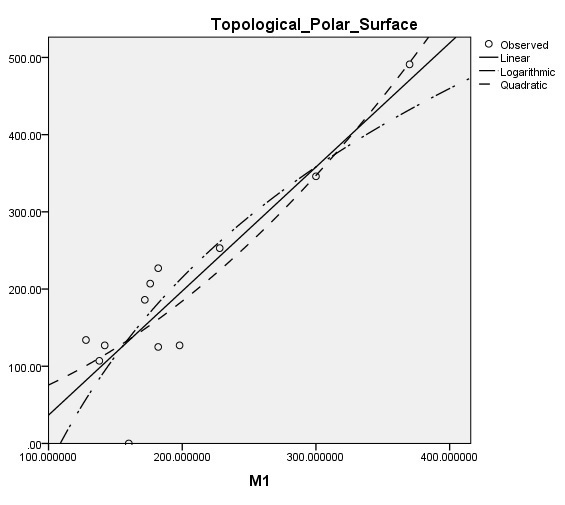}
\includegraphics[width = 3.0in]{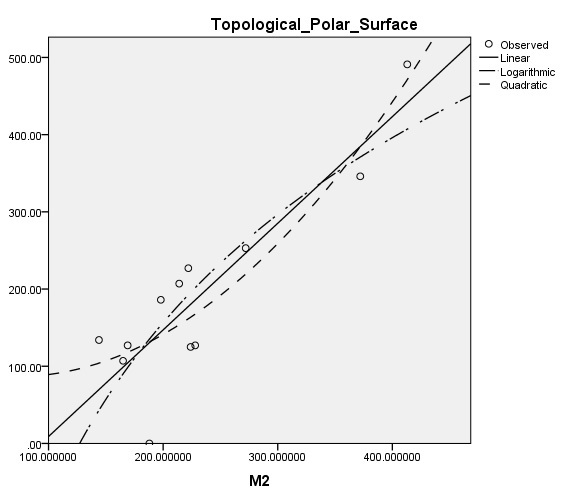}
\end{center}
\begin{center}
\includegraphics[width = 3.0in]{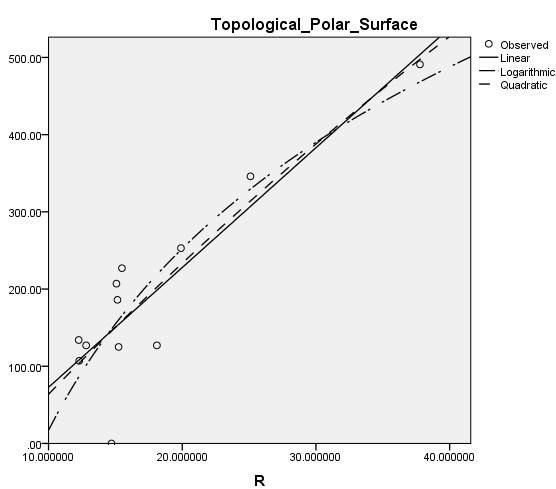}
\includegraphics[width = 3.0in]{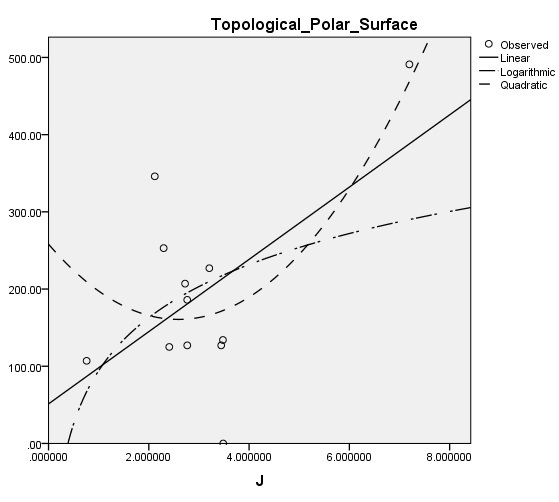}
\end{center}
\begin{center}
\includegraphics[width = 3.0in]{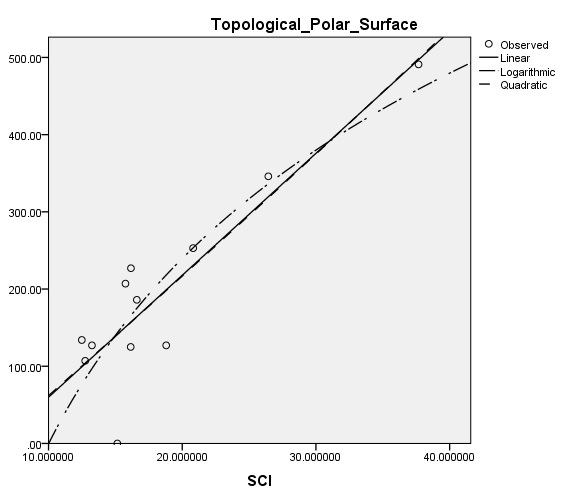}
\end{center}

\begin{center}
\textbf{Table 6.} Correlation coefficient, $F$ and $S$ values. \\
\begin{tabular}{|c|c|c|c|}
  \hline
  Model No & $R^{2}$ & $F$ & $S$ \\
\hline
Model 51&0.804	&41.023	&0.000\\
Model 56& 0.753	&30.540	&0.000\\
Model 61& 0.815	&19.793		&0.001\\
Model 52& 0.787	&36.898			&0.000\\
Model 57& 0.726	&26.456			&0.000\\
Model 62& 0.811	&19.311			&0.001\\
Model 53& 0.780	&35.410			&0.000\\
Model 58& 0.761	&31.783			&0.000\\
Model 63& 0.781	&16.030			&0.001\\
Model 54& 0.774	&34.248			&0.000\\
Model 59& 0.749	&29.906			&0.000\\
Model 64& 0.774	&15.412			&0.001\\
Model 55& 0.781	&35.596			&0.000\\
Model 60& 0.748	&29.610			&0.000\\
Model 65& 0.782	&16.102			&0.001\\
\hline
\end{tabular}
\end{center}
By looking at the above table we can see that the predicting power of above mentioned topological indices are better with respect to topological polar surface area of phytochemicals. The correlation coefficient of the first Zagreb index ($M_{1}$) lies between 0.753 to 0.815, whereas, the range for the Zagreb index($M_{2}$) is lies between 0.726 to 0.811. For Randi$\acute{c}$($R$) index the $r$ values is lies between 0.761 to 0.781 and for the Balban index $J$ the $r$ value ranging from 0.749 to 0.774. Finally for sum-connectivity index, the $r$ value lies between 0.748 to 0.782.  Among all $TI's$, the first Zagreb index $M_{1}$ has better predicting power than other topological indices with respect to  topological polar surface area of phytochemicals.
\vspace{2mm}

\noindent \textbf{Conclusion:} The QSPR study has revealed that the molecular descriptors are best candidates to predict the physicochemical properties of phyctochemicals. In particular, the sum-connectivity index $(SCI)$ and the first Zagreb index $(M_{1})$ are two important parameters to predict the molecular weight and the topological polar surface area of phytochemicals respectively. Our study may help the researchers in the field of life-science in finding the anti-covid-19 drugs.

\end{document}